\documentclass[letterpaper, 10 pt, conference]{ieeeconf}  

\usepackage{amssymb}
\usepackage{amstext}
\usepackage{amsmath}
\usepackage{epsfig}
\usepackage{epstopdf}
\usepackage{color}
\usepackage{graphicx}
\usepackage{booktabs}

\IEEEoverridecommandlockouts                              
\title{\LARGE \bf
Lyapunov Method Based Online Identification of Nonlinear Systems Using Extreme Learning Machines
}

\author{Vijay Manikandan Janakiraman$^{1}$ and Dennis Assanis$^{2}$
\thanks{*This work was not supported by any organization}
\thanks{$^{1}$Vijay Manikandan is a PhD Candidate, Mechanical Engineering,
        University Michigan, Ann Arbor. MI, USA
        {\tt\small vijai at umich.edu}}%
\thanks{$^{2}$D. Assanis is with the Stony Brook University,
        New York, USA
        {\tt\small dennis.assanis@stonybrook.edu}}%
}

\begin{document}

\maketitle
\thispagestyle{empty}
\pagestyle{empty}

\begin{abstract}

Extreme Learning Machine (ELM) is an emerging learning paradigm for nonlinear regression problems and has shown its effectiveness in the machine learning community. An important feature of ELM is that the learning speed is extremely fast thanks to its random projection preprocessing step. This feature is taken advantage of in designing an online parameter estimation algorithm for nonlinear dynamic systems in this paper. The ELM type random projection and a nonlinear transformation in the hidden layer and a linear output layer is considered as a generalized model structure for a given nonlinear system and a parameter update law is constructed based on Lyapunov principles. Simulation results on a DC motor and Lorentz oscillator show that the proposed algorithm is stable and has improved performance over the online-learning ELM algorithm.
\end{abstract}

\section{INTRODUCTION}
System identification is the process of obtaining mathematical models of systems using input-output data. System identification is important in design and analysis of control systems when the development of a physics-based dynamical model is not trivial. Several algorithms for identification of a linear system exist \cite{Ljung,nelles} but when the nonlinearity is of a higher order, the local linear assumption fails and it becomes important to develop nonlinear identification methods. There exist online identification algorithms for nonlinear systems as well. Since the underlying structure is not assumed for the nonlinear system, a neural network type model can be a good choice \cite{narendra,ioannou} among others. Such algorithms rely on linearizing the basis functions to obtain the gradient of the output error with respect to the network parameters. Different from the previous approaches, this paper makes use of the recently developed Extreme Learning Machines (ELM) for mapping the system nonlinearity. By exploiting ELM's random projection preprocessing stage where the input data is projected onto a high dimensional space where the features can be mapped using a linear least squares method, the high speed learning of ELM is inherited in the proposed algorithm. Using a Lyapunov method, a stable parameter update law for nonlinear system identification has been developed for continuous time dynamic systems.

\section{\uppercase{Extreme Learning Machines - A Review}} \label{elm review}
Extreme Learning Machine (ELM) is an emerging learning paradigm for multi-class classification and regression problems \cite{4Huang2005,huang12}. The highlight of ELM compared to the other state of the art methodologies like neural networks, support vector machines is that the training speed of ELM is extremely fast. The key enabler for ELM's training speed is the random assignment of input layer parameters which do not require adaptation to the data. In such a setup, the output layer parameters can be determined analytically using least squares. Some of the attractive features of ELM \cite{4Huang2005} are listed below

\begin{enumerate}
  \item ELM is an universal approximator
  \item ELM results in the smallest training error without getting trapped in local minima (better accuracy)
  \item ELM does not require iterative training (low computational demand)
  \item ELM solution has the smallest norm of weights (better generalization)
  \item The minimum norm least square solution by ELM is unique.
\end{enumerate}

\begin{figure}[]
      \centering
      \includegraphics[scale=0.5]{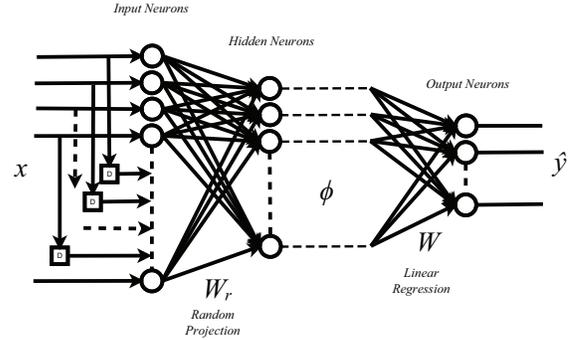}
      \caption{ELM Model Structure.}
      \label{elmfig}
\end{figure}

ELM is developed from a machine learning perspective and hence data observations are considered independent and identically distributed. Hence the observations are discrete and a dynamic system application may not be directly suitable as the data is connected in time. However, ELM can be applied for system identification in discrete time by using a series-parallel formulation \cite{narendra}. A generic nonlinear identification using the nonlinear auto regressive model with exogenous input (NARX) is considered as follows
\begin{equation}\label{NARX eqn}
y(k)=f[u(k-1),..,u(k-n_u),y(k-1),..,y(k-n_y)]
\end{equation}
where $u(k)\in\mathbb{R}^{u_d}$ and $y(k)\in\mathbb{R}^{y_d}$ represent the inputs and outputs of the system respectively, $k$ represents the discrete time index, $f(.)$ represents the nonlinear function mapping specified by the model, $n_u$, $n_y$ represent the number of past input and output samples required (order of the system) for prediction while $u_d$ and $y_d$ represent the dimension of inputs and outputs respectively.

\subsection{Offline learning algorithm}
The input-output measurement sequence of system \eqref{NARX eqn} can be converted to the form of training data as required by ELM
\begin{equation}\label{}
\{(x_1,y_1),...,(x_N,y_N)\}\in \big(\mathcal{X},\mathcal{Y}\big)
\end{equation}
where $\mathcal{X}$ denotes the space of the input features (Here $\mathcal{X} = \mathbb{R}^{u_dn_u+y_dn_y}$ and $\mathcal{Y} = \mathbb{R}^{y_d}$) and $x$ represent the augmented input vector obtained by appending the input and output measurements from the system as follows
\begin{equation}\label{inp feature}
x=[u(k-1),..,u(k-n_u),y(k-1),..,y(k-n_y)]^T
\end{equation}

The ELM is an unified representation of single layer feed-forward networks (SLFN) and is given by \eqref{elm model} where $g$ represents the hidden layer activation function and $W_r, W$ represents the input and output layer parameters respectively.
\begin{equation}\label{elm model}
\hat{y}=[g(W_r^Tx+b_r)]^TW
\end{equation}
The matrix $W_r$ consists of randomly assigned elements that maps the input vector to a high dimensional feature space while $b_r$ is a bias component assigned in a random manner similar to $W_r$. The elements can be assigned based on any continuous random distribution \cite{huang12} and remains fixed during training. The number of hidden neurons determine the dimension of the transformed feature space and the hidden layer is equipped with a nonlinear activation function similar to traditional neural network architecture. It should be noted that nonlinear regression using neural networks for instance, the input layer parameters $W_r$ and $W$ are simultaneously adjusted during training. Since there is a nonlinear connection between the two layers, iterative techniques are the only possible solution. ELM, however, avoids the iterative training as the input layer parameters are randomly selected \cite{4Huang2005}. Hence the training step of ELM reduces to finding a least squares solution to the output layer parameters $W$ given by
\begin{equation}\label{}
\min_{W}\left\{\|HW-Y\|^2+\lambda \|W\|^2\right\}
\end{equation}
\begin{equation}\label{}
\hat{W}=\left(\frac{I}{\lambda}+H^TH \right)^{-1}H^TY
\end{equation}
where $\lambda$ represents the regularization coefficient, T represents the vector of outputs or targets and $H$ the hidden layer output matrix as termed in literature (see Figure \ref{elmfig}).

\subsection{Online learning algorithm}
In the batch training mode (offline training), all the data is assumed to be present. However, for an online system identification problem, data is sampled continuously and is available one by one. Hence the sequential learning algorithm can be modified to perform identification. The ELM online sequential algorithm can be formulated as follows \cite{oselm}

As an initialization step, a set of data observations are required to initialize the $H_0$ and $W_0$ by solving
\begin{equation}\label{}
\min_{W_0}\left\{\|H_0W_0-Y_0\|^2+\lambda \|W_0\|^2\right\}
\end{equation}
\begin{equation}\label{}
H_0 = (W_r^T x_0+b_0)^T \in \mathbb{R}^{n_0 \times n_h}
\end{equation}
where $n_0$ and $n_h$ represents the number of data observations in the initialization step and the number of hidden neurons of the ELM model respectively. The solution $W_0$ is given by
\begin{equation}\label{}
W_0=K_0^{-1} H_0^T Y_0
\end{equation}
where $K_0 = H_0^T H_0$. Suppose given another new data $x_1$, the problem becomes
\begin{equation}\label{}
\min_{W_1}\left\|
\left[\begin{array}{cc}
H_0 \\
H_1 \end{array}\right]
W_1 -
\left[\begin{array}{cc}
Y_0\\
Y_1 \end{array}\right]
\right\|^2
\end{equation}
The solution can be derived as
\begin{eqnarray*}
  W_1 &=& W_0 + K_1^{-1} H_1^T (T_1 - H_1 W_0) \\
  K_1 &=& K_0 + H_1^T H_1
\end{eqnarray*}
Based on the above, a generalized recursive algorithm for updating the least-squares solution can be computed as follows
\begin{equation}
  P_{k+1} = P_k - P_k H^T_{k+1} (I + H_{k+1} P_k H_{K+1}^T )^{-1} H_{k+1} P_k
\end{equation}
\begin{equation}
  W_{k+1} = W_k + P_{k+1} H^T_{k+1} (T_{k+1} - H_{k+1} W_k)
\end{equation}

\section{\uppercase{Lyapunov based parameter Update law}}
The parameter update law is derived for a continuous time system. A general multi-input multi-output (MIMO) nonlinear dynamic system is given by
\begin{equation}\label{gen_sys}
\dot{z}(t)=f(z(t),u(t))
\end{equation}
where the state vector $z\in\mathbb{R}^{n\times1}$, input (or control) vector $u\in\mathbb{R}^{m\times1}$. By adding and subtracting $Az(t)$ where $A\in\mathbb{R}^{n\times n}$ is a Hurwitz matrix, then the system \eqref{gen_sys} becomes
\begin{equation}\label{con_sys}
\dot{z}(t)=Az(t)+g(z(t),u(t))
\end{equation}
where $g(z(t),u(t))=f(z(t),u(t))-Az(t)$ describes the system nonlinearity. Assuming ELM can model the system nonlinearity $g(z(t),u(t))$ with an accuracy of $\epsilon$. If we assume bounded inputs and bounded states for the system \eqref{gen_sys}, then $\epsilon(t)$ for the model is finite and is bounded above by $\xi$ \cite{4Huang2005}. The system \eqref{con_sys} can now be represented by
\begin{equation}\label{elm_sys}
\dot{z}(t)=Az(t)+W^T_*\phi+\epsilon(t)
\end{equation}
The parametric model of the system can be considered as
\begin{equation}\label{par_sys}
\dot{\hat{z}}(t)=A\hat{z}(t)+\hat{W}^T\phi
\end{equation}
where $W_*$ and $\hat{W}$ represents the actual and estimated parameters of the ELM model, $\phi$ represents the hidden layer output of ELM (see Figure \ref{elmfig}). It should be noted that the input-hidden layer connection parameters $W_r$ has been chosen randomly and fixed assuming that ELM only needs tuning of the output layer weights $W$. Hence $\phi$ can be considered the same for both the system and the parametric model which is a simplification that has been achieved with the help of the ELM formulation. This simplicity cannot be achieved using traditional back-propagation neural networks and hence the strength of the proposed method. The estimation error and the error dynamics are given by
\begin{equation}\label{est_err}
e(t)=z-\hat{z}
\end{equation}
\begin{eqnarray}\label{err_dyn}
  \dot{e}(t) &=& A e(t)+(W^T_*-\hat{W}^T)\phi+\epsilon(t) \\
    &=& A e(t)+\tilde{W}^T\phi+\epsilon(t)
\end{eqnarray}
where $\tilde{W}$ represents the parameter error.

In order to have a stable parameter update law that guarantees convergence of both estimation error and the parametric error to zero, the following Lyapunov function is considered.

\begin{equation}\label{lyap_fn}
V=\frac{1}{2} e^Te + \frac{1}{2} tr(\tilde{W}^T\tilde{W})
\end{equation}
\begin{eqnarray}\label{}
\nonumber\dot{V} &=& e^T\dot{e} + tr(\tilde{W}^T\dot{\tilde{W}}) \\
\nonumber&=& e^TAe + e^T\tilde{W}^T\phi + e^T\epsilon(t) + tr(\tilde{W}^T\dot{\tilde{W}}) \\
\nonumber&=& e^TAe + e^T\tilde{W}^T\phi + e^T\epsilon(t) + \sum_{i=1}^{n} \tilde{w_i}^T\dot{\tilde{w_i}} \\
\nonumber&=& e^TAe + e^T\epsilon(t) + \sum_{i=1}^{n} \phi^T\tilde{w_i}e_i + \sum_{i=1}^{n} \nonumber\tilde{w_i}^T\dot{\tilde{w_i}}
\end{eqnarray}

if we choose $\dot{\tilde{w_i}}$ such that
\begin{eqnarray}\label{}
\nonumber\tilde{w_i}^T\dot{\tilde{w_i}} &=& -\phi^T\tilde{w_i}e_i \\
\nonumber\dot{\tilde{w_i}}^T\tilde{w_i} &=& -\phi^T\tilde{w_i}e_i \\
\nonumber\dot{\tilde{w_i}}^T &=& -\phi^Te_i \\
\nonumber\dot{\tilde{w_i}} &=& -\phi e_i \\
\dot{\hat{w_i}} &=& \phi e_i
\end{eqnarray}
then $\dot{V}$ becomes
\begin{eqnarray}\label{}
\nonumber\dot{V} &=& e^TAe + e^T\epsilon(t) \\
 & \leq & \|e\|_2 |\lambda_{max}(A)| \|e\|_2 + \xi\|e\|_2
\end{eqnarray}
However, $\dot{V}<0$ if
\begin{equation}\label{lyap_condition}
\|e\|_2>\frac{\xi}{|\lambda_{max}(A)|}=\Gamma
\end{equation}
By applying the universal approximation capability of ELM, the approximation error $\epsilon$ can be made arbitrarily small and hence $\Gamma$ converges to zero. Hence with proper selection of the number of hidden neurons $n_h$ of ELM and with persistent excitation, both the estimation error $e$ as well as the parameter error $\tilde{W}$ can be made to converge to zero. It should be noted that as long as the estimation error is above $\Gamma$, the stability of the algorithm is guaranteed. The value of $\Gamma$ can be chosen to be the required accuracy of ELM approximation \cite{Yan2000,ioannou} so that the adaptation can occur as long as the model approximation error is greater than the required accuracy.

Hence the parameter estimation algorithm based on Lyapunov analysis is given by
\begin{eqnarray}\label{lyap_law}
\dot{\hat{W}} &=& \phi e^T
\end{eqnarray}

\section{\uppercase{Simulations}}
The two algorithms compared for the simulation study are the existing online ELM algorithm \cite{oselm} and the proposed Lyapunov based ELM algorithm. For all the simulations, the same ELM model structure with the same randomly assigned input layer weights and biases ($W_r$ and $b_r$) as well as the same initial condition for output layer weights ($W_0$) are imposed. The design matrix $A$ can also be appropriately chosen so as to suit the requirements on overshoot, settling time of the parameter estimation \cite{Yan2000,ioannou}.

It should be noted that the input layer parameters $W_r$ is fixed. It is required by ELM that all data is normalized to lie between -1 and +1 and hence appropriate scaling in introduced during simulation. The limits of the states and inputs are known a priori and can be used in the normalization. The inputs to the system has to be persistently exciting (as required for parameter convergence) which not easy to achieve in nonlinear systems. Hence the input signal follows a pseudo-random multi level sequence (PRMS) which represents several combination of step inputs at different magnitudes and frequencies suitable for exciting nonlinear systems \cite{prbs_sysid}.

\subsection{DC motor example}
A nonlinear DC motor system is considered whose dynamic equations are as follows

\begin{eqnarray}
  \dot{x} &=& f(x) + g(x)u
\end{eqnarray}
where
\[ f(x)= \left [ \begin{array}{cc}
-c_1x_1 + c_3  \\
-c_4x_2  \end{array} \right ] \]
\[ g(x)= \left [ \begin{array}{cc}
-c_2x_2  \\
-c_5x_1  \end{array} \right ] \]
where $c_1$=60, $c_2$=0.5, $c_3$=40, $c_4$=6, $c_5$=40000.
The design matrix $A$ is chosen as
\[ A = \left [ \begin{array}{cc}
-50 & 0  \\
0 & -50  \end{array} \right ] \]

\begin{figure}[]
      \centering
      \includegraphics[scale=0.6]{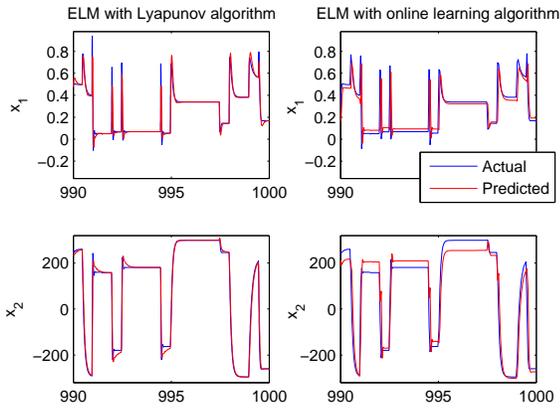}
      \caption{Comparison of system states of actual and estimated system by Lyapunov ELM and Online ELM for DC motor system.}
      \label{dc_motor_states}
\end{figure}

\begin{figure}[]
      \centering
      \includegraphics[scale=0.6]{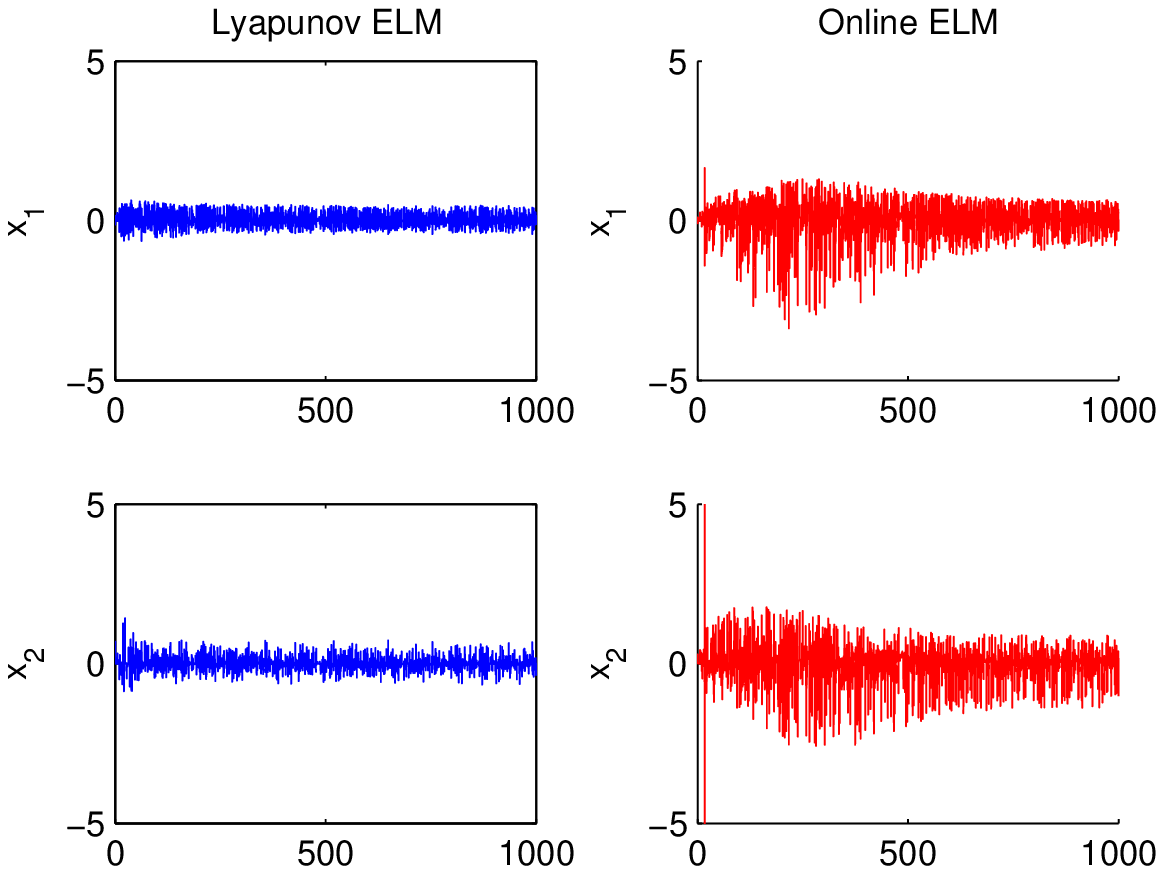}
      \caption{Convergence of error between the states of actual and estimated system by Lyapunov ELM and Online ELM for DC motor system.}
      \label{dc_motor_err}
\end{figure}

\begin{figure}[]
      \centering
      \includegraphics[scale=0.6]{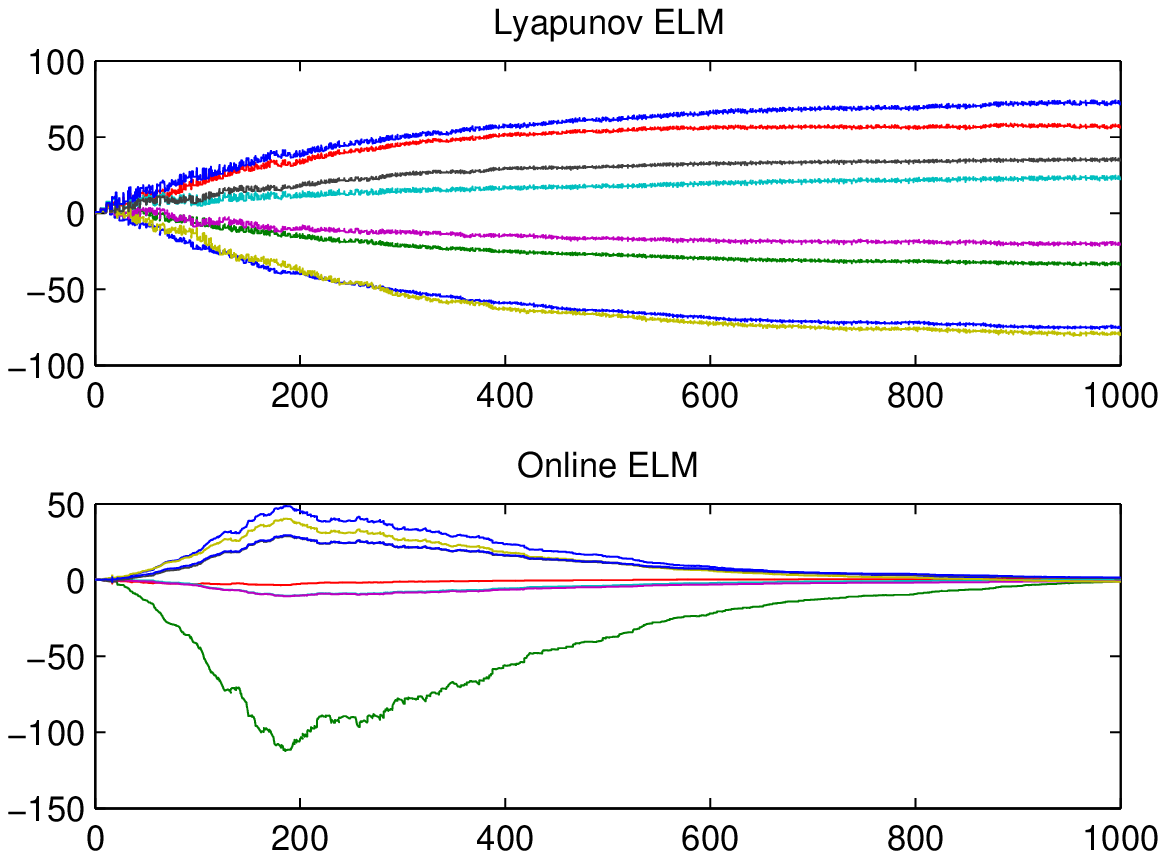}
      \caption{Parametric Convergence (only few parameters shown) by Lyapunov ELM and Online ELM for DC motor system.}
      \label{dc_motor_par}
\end{figure}

\begin{figure}[]
      \centering
      \includegraphics[scale=0.6]{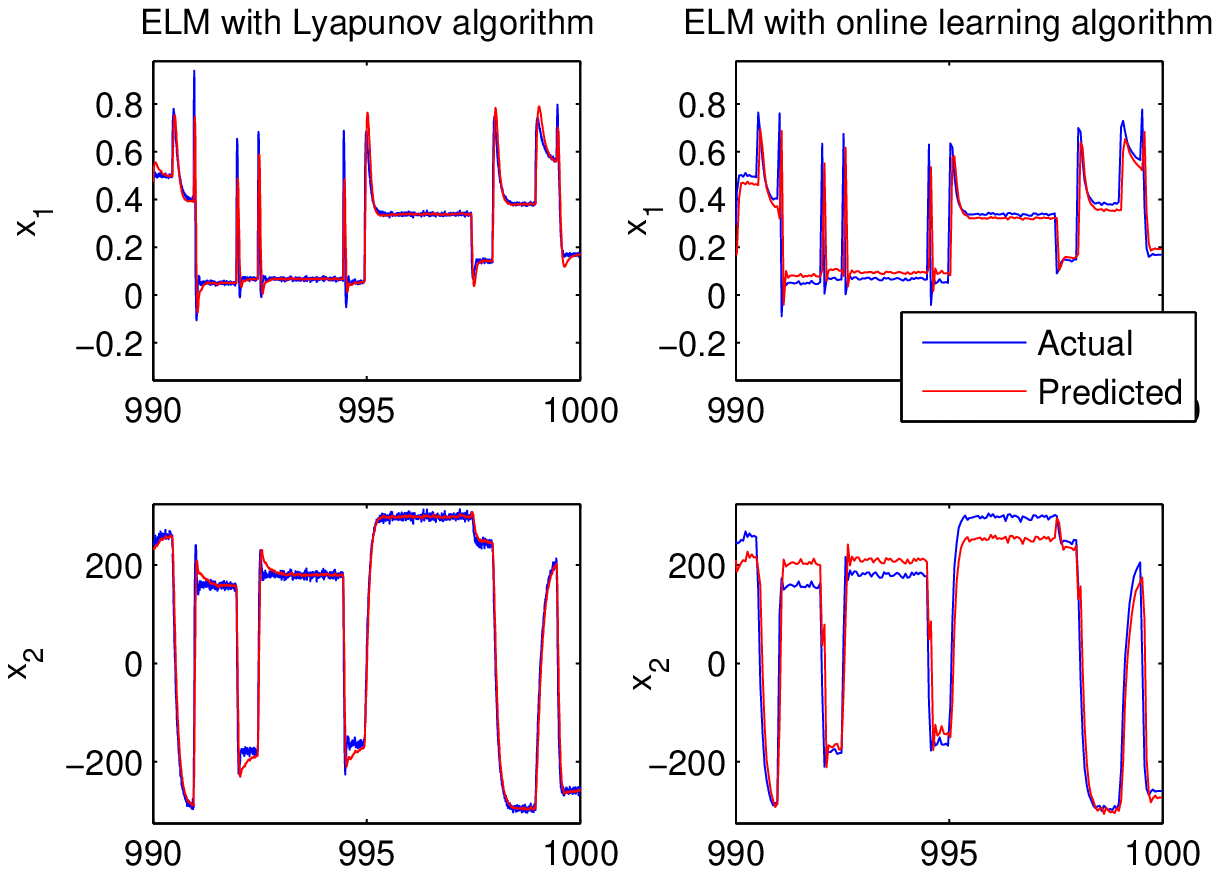}
      \caption{Comparison of system states of actual and estimated system by Lyapunov ELM and Online ELM for DC motor system with gaussian measurement noise.}
      \label{dc_motor_states_noise}
\end{figure}

\begin{figure}[]
      \centering
      \includegraphics[scale=0.6]{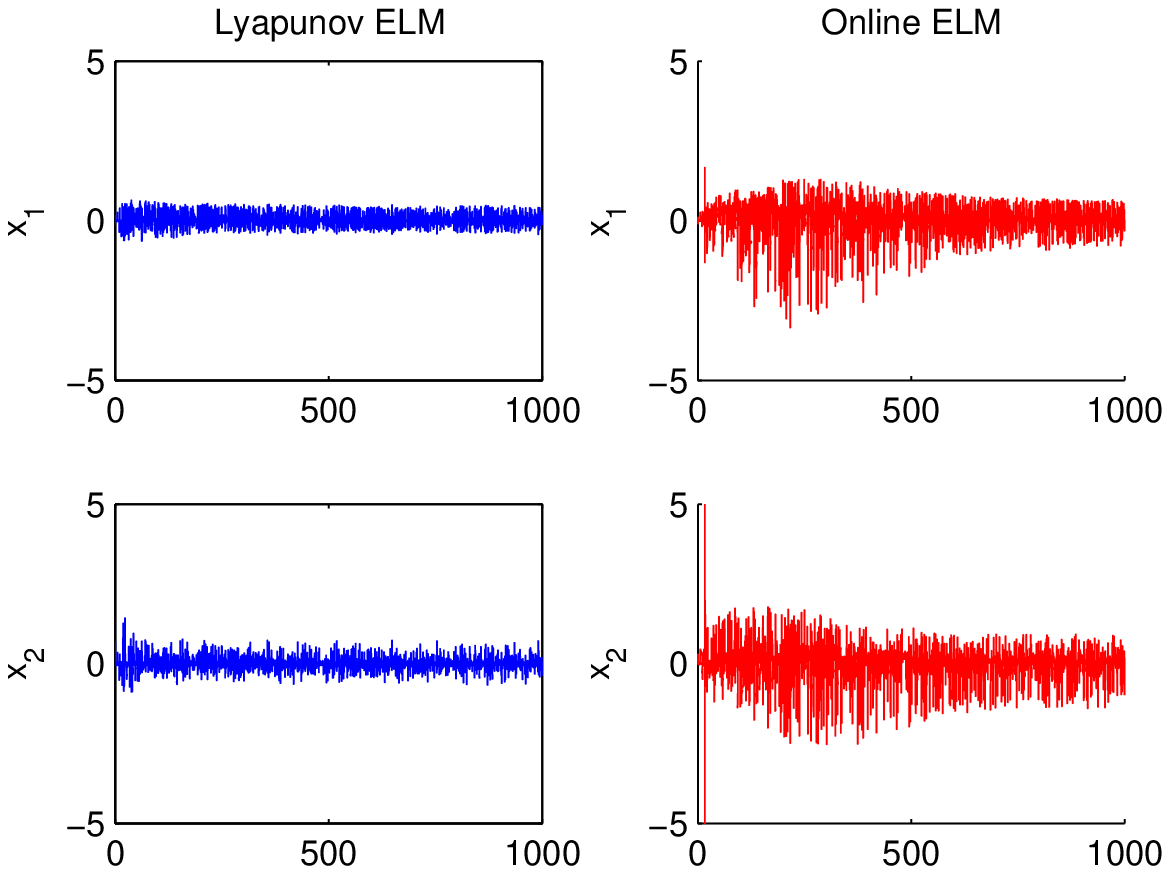}
      \caption{Convergence of error between the states of actual and estimated system by Lyapunov ELM and Online ELM for DC motor system with gaussian measurement noise.}
      \label{dc_motor_err_noise}
\end{figure}

\begin{figure}[]
      \centering
      \includegraphics[scale=0.6]{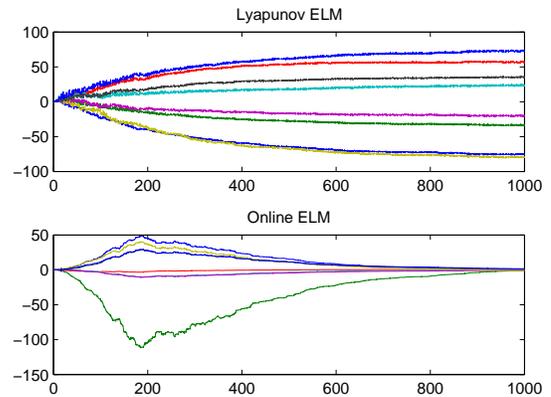}
      \caption{Parametric Convergence (only few parameters shown) by Lyapunov ELM and Online ELM for DC motor system with gaussian measurement noise.}
      \label{dc_motor_par_noise}
\end{figure}

The number of hidden neurons for ELM model is chosen as 8. Sigmoidal activation function is considered as the input layer activation function. Two cases are compared - with and without gaussian noise at the measurement. The results are summarized in Figures \ref{dc_motor_states}-\ref{dc_motor_par} for the case without noise and in Figures \ref{dc_motor_states_noise}-\ref{dc_motor_par_noise} for the case with noise. The results of root mean squared error (RMSE) between the states of the actual and estimated system are compared in Table \ref{dc_results}.

\begin{table}[]
  \centering
  \caption{Comparison of normalized RMSE of the error between the states of the nonlinear system and the models by Online ELM and Lyapunov ELM for the DC motor system}
    \begin{tabular}{ccc}
    \toprule
          & Online ELM & Lyapunov ELM \\
    \midrule
    normalized RMSE & 0.4635 & 0.0935 \\
    normalized RMSE (with noise) & 0.4626 & 0.0936 \\
    \bottomrule
    \end{tabular}%
  \label{dc_results}%
\end{table}%

\subsection{Lorentz oscillator}
A chaotic dynamic system is a nonlinear deterministic system that displays nonlinear and unpredictable behavior. These systems are very sensitive to initial conditions and system's parameters behavior. One of the ways to represent a chaotic system is using Lorentz system whose dynamic equations are as follows
\begin{eqnarray*}
\dot{x} &=& \sigma(y-x) \\
\dot{y} &=& rx-y-xz \\
\dot{z} &=& xy-bz
\end{eqnarray*}
where $\sigma, r, b > 0$ are system parameters. For this simulation, $\sigma$=10, $r$=28 and $b$=8/3 are considered. It should be noted that there are no excitation input to the system.

The design matrix $A$ is chosen as
\[ A = \left [ \begin{array}{ccc}
-60 & 0 & 0 \\
0 & -60 & 0 \\
0 & 0 & -120
\end{array} \right ] \]

\begin{figure}[]
      \centering
      \includegraphics[scale=0.6]{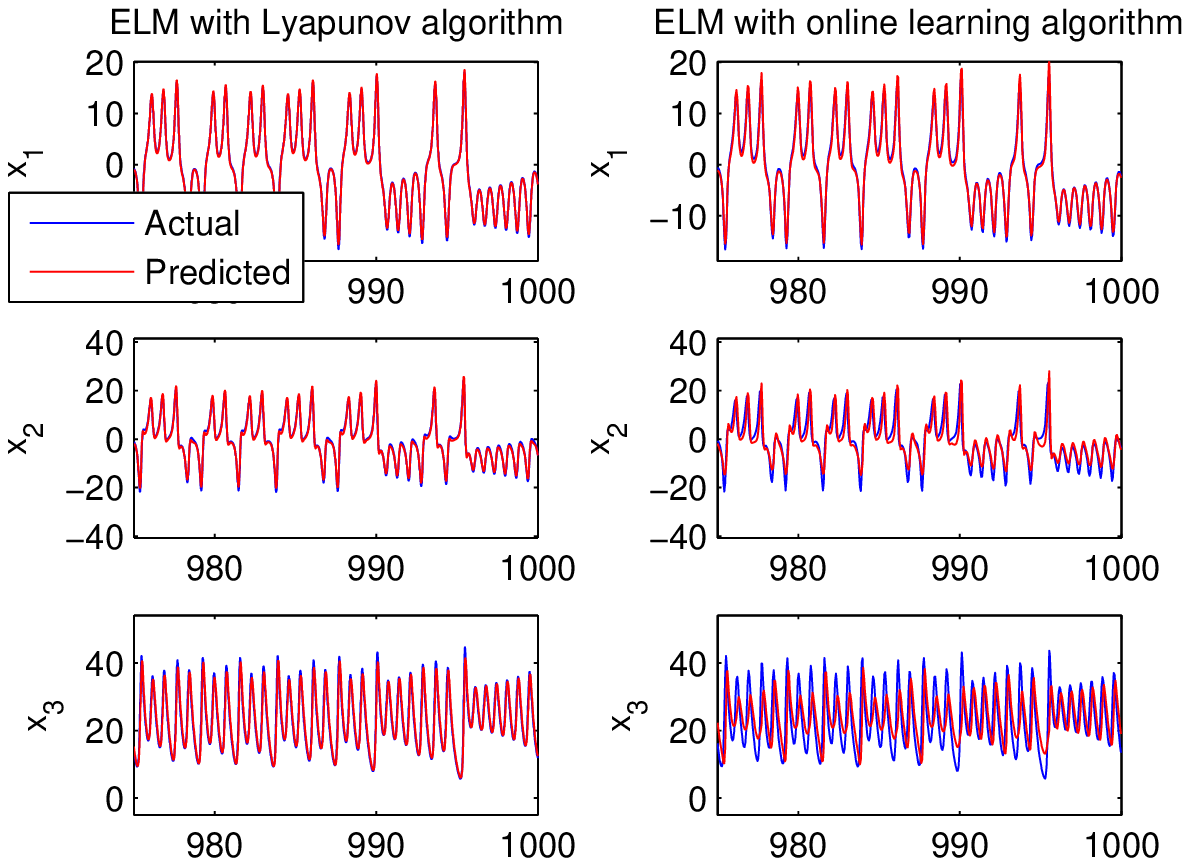}
      \caption{Comparison of system states of actual and estimated system by Lyapunov ELM and Online ELM for Lorentz system.}
      \label{lor_states}
\end{figure}

\begin{figure}[]
      \centering
      \includegraphics[scale=0.6]{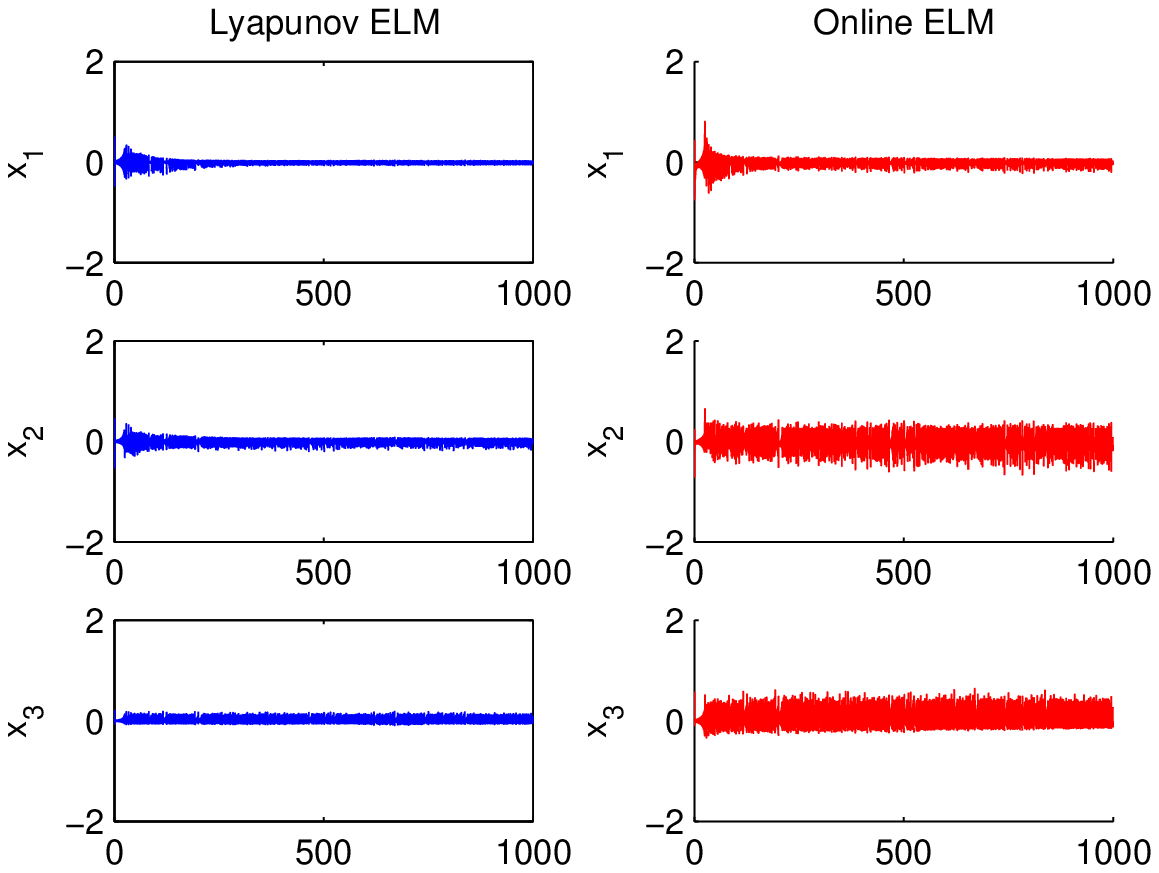}
      \caption{Convergence of error between the states of actual and estimated system by Lyapunov ELM and Online ELM for Lorentz system.}
      \label{lor_err}
\end{figure}

\begin{figure}[]
      \centering
      \includegraphics[scale=0.6]{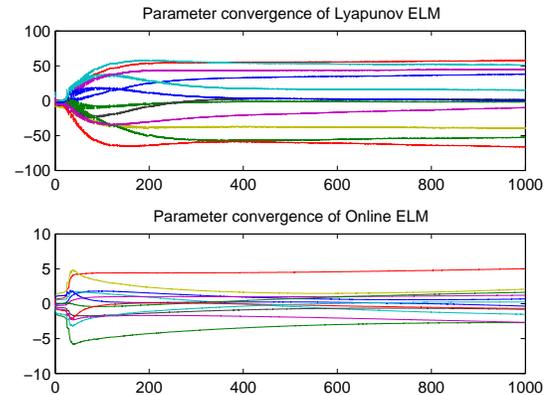}
      \caption{Parametric Convergence (only few parameters shown) by Lyapunov ELM and Online ELM for Lorentz system.}
      \label{lor_par}
\end{figure}

\begin{figure}[]
      \centering
      \includegraphics[scale=0.6]{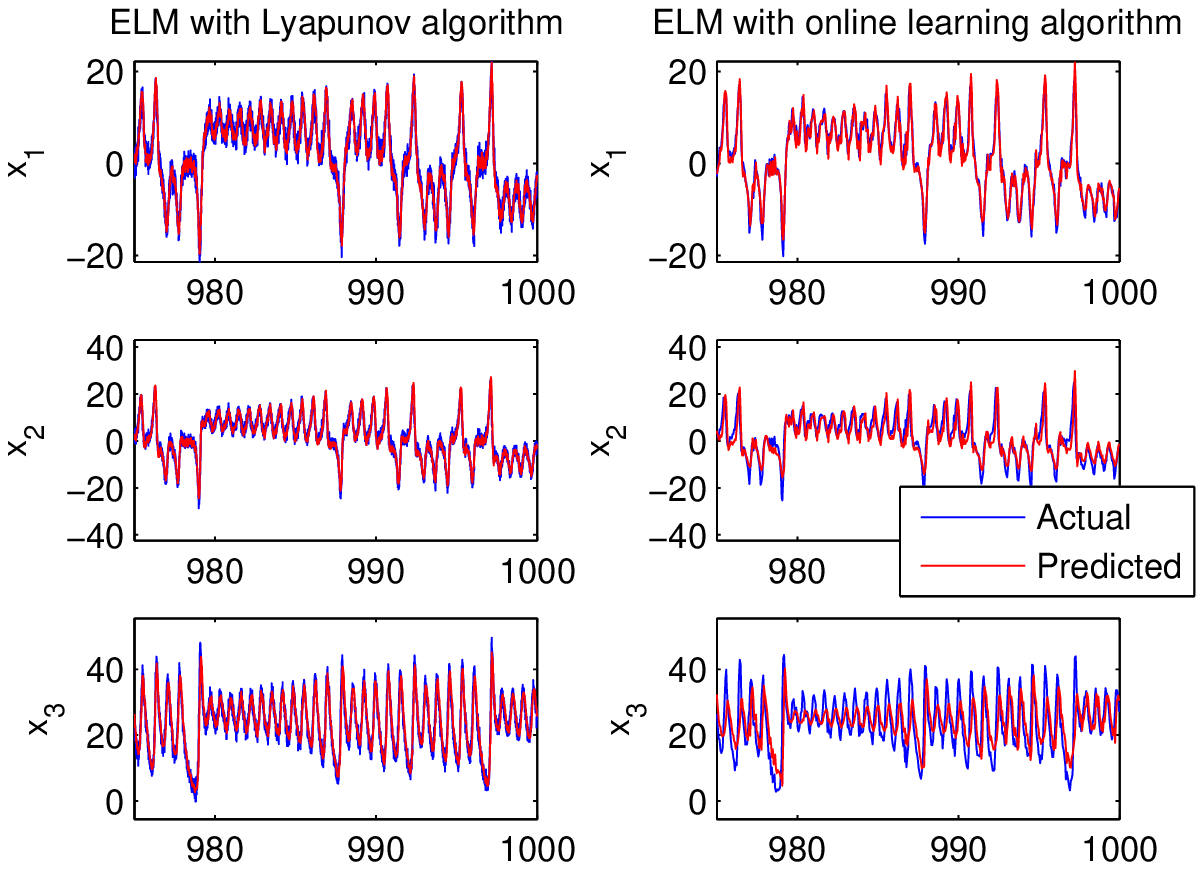}
      \caption{Comparison of system states of actual and estimated system by Lyapunov ELM and Online ELM for Lorentz system with gaussian measurement noise.}
      \label{lor_states_noise}
\end{figure}

\begin{figure}[]
      \centering
      \includegraphics[scale=0.6]{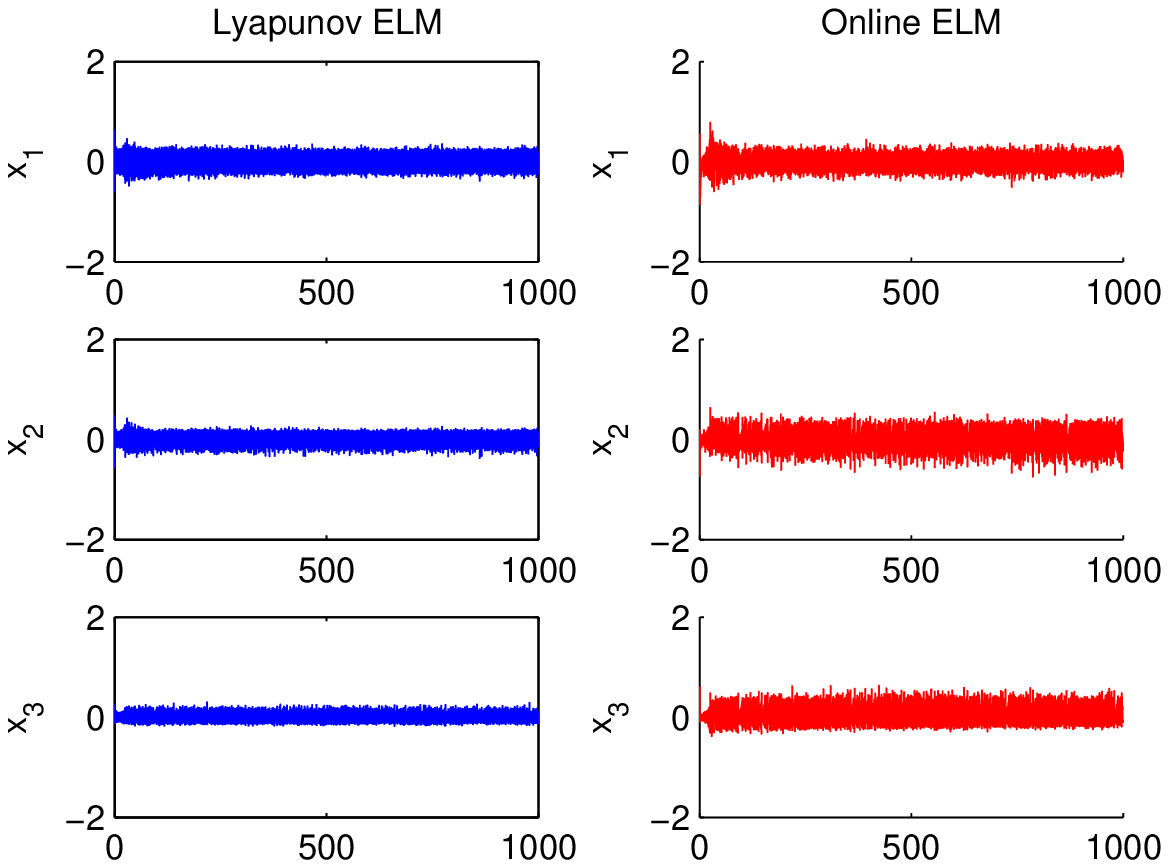}
      \caption{Convergence of error between the states of actual and estimated system by Lyapunov ELM and Online ELM for Lorentz system with gaussian measurement noise.}
      \label{lor_err_noise}
\end{figure}

\begin{figure}[]
      \centering
      \includegraphics[scale=0.6]{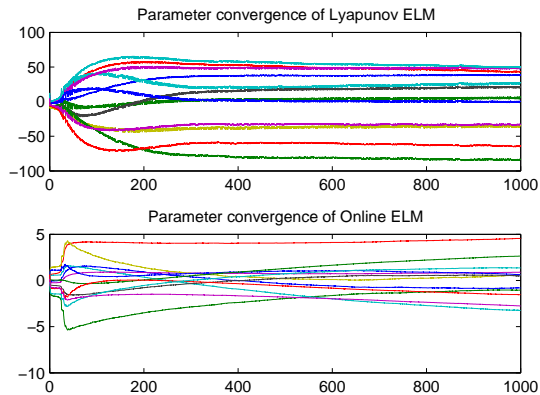}
      \caption{Parametric Convergence (few parameters shown) by Lyapunov ELM and Online ELM for Lorentz system with gaussian measurement noise.}
      \label{lor_par_noise}
\end{figure}

The number of hidden neurons for ELM model is chosen as 12. Sigmoidal activation function is considered as the input layer activation function. Two cases are compared - with and without gaussian noise at the measurement. The results are summarized in Figures \ref{lor_states}-\ref{lor_par} for the case without noise and in Figures \ref{lor_states_noise}-\ref{lor_par_noise} for the case with noise. The results of root mean squared error between the states of the actual and estimated system are compared in Table \ref{lor_results}.

\begin{table}[]
  \centering
  \caption{Comparison of normalized RMSE of the error between the states of the nonlinear system and the models by Online ELM and Lyapunov ELM for the Lorentz system.}
    \begin{tabular}{ccc}
    \toprule
          & Online ELM & Lyapunov ELM \\
    \midrule
    normalized RMSE  & 0.2085 & 0.0652 \\
    normalized RMSE (with noise) & 0.2424 & 0.1139 \\
    \bottomrule
    \end{tabular}%
  \label{lor_results}%
\end{table}%

\section{Discussion}
It can be observed from the simulation results that the proposed Lyapunov ELM algorithm is suited for nonlinear system identification and has performance better than a sequential learning online ELM algorithm. From Figures \ref{dc_motor_err} and \ref{lor_err}, it can be observed that the states of the system and the estimated model converge for both examples. From Figures \ref{dc_motor_par}, \ref{lor_par}, the convergence of model parameters can be seen but it is not guaranteed that the parameters converge to their true values as the model structure takes a general form and independent of the actual system. The above are observed for the cases with measurement noise too. It can also be observed from Figures \ref{dc_motor_par} and Figures \ref{dc_motor_par_noise} that parameter convergence may be faster for the Lyapunov ELM case compared to the online ELM algorithm. Also, parameter convergence appears to be monotonic for the Lyapunov ELM case. Finally, from Tables \ref{dc_results} and \ref{lor_results}, it can be observed that the Lyapunov ELM outperforms online ELM algorithm and achieves a better accuracy in terms of the estimated states. It should be noted that the design matrix $A$ needs tuning depending on the nature of transient response in prediction. However it is straightforward as an decrease in the magnitude of the eigen values of A results in a faster tracking. This gives additional flexibility and control on the Lyapunov ELM's performance. 

\section{CONCLUSIONS}

An online system identification algorithm for nonlinear systems has been developed using a Lyapunov approach. The complexity of the proposed algorithm is similar to that of a linear parameter estimation thanks to the random preprocessing step featured by extreme learning machines. The proposed algorithm carries over the simplicity of ELM but performs better than the online version of ELM owing to the stability guarantee of Lyapunov's method. Simulation results on two examples prove the validity of the proposed algorithm. Future Work will focus on application to a complex real world nonlinear dynamic system and study convergence properties.

\addtolength{\textheight}{-12cm}   

\bibliographystyle{IEEEtran}
\bibliography{IEEEabrv,ELM_sysid}

\begin{thebibliography}{10}
\providecommand{\url}[1]{#1}
\csname url@rmstyle\endcsname
\providecommand{\newblock}{\relax}
\providecommand{\bibinfo}[2]{#2}
\providecommand\BIBentrySTDinterwordspacing{\spaceskip=0pt\relax}
\providecommand\BIBentryALTinterwordstretchfactor{4}
\providecommand\BIBentryALTinterwordspacing{\spaceskip=\fontdimen2\font plus
\BIBentryALTinterwordstretchfactor\fontdimen3\font minus
  \fontdimen4\font\relax}
\providecommand\BIBforeignlanguage[2]{{%
\expandafter\ifx\csname l@#1\endcsname\relax
\typeout{** WARNING: IEEEtran.bst: No hyphenation pattern has been}%
\typeout{** loaded for the language `#1'. Using the pattern for}%
\typeout{** the default language instead.}%
\else
\language=\csname l@#1\endcsname
\fi
#2}}

\bibitem{Ljung}
L.~Ljung, Ed., \emph{System identification (2nd ed.): theory for the
  user}.\hskip 1em plus 0.5em minus 0.4em\relax Upper Saddle River, NJ, USA:
  Prentice Hall PTR, 1999.

\bibitem{nelles}
O.~Nelles, \emph{{Nonlinear System Identification: From Classical Approaches to
  Neural Networks and Fuzzy Models}}, 1st~ed.\hskip 1em plus 0.5em minus
  0.4em\relax Springer, Dec. 2000.

\bibitem{narendra}
\BIBentryALTinterwordspacing
K.~S. Narendra and K.~Parthasarathy, ``Identification and control of dynamical
  systems using neural networks,'' \emph{{IEEE} Trans. Neural Networks},
  vol.~1, no.~1, pp. 4--27, Mar. 1990. [Online]. Available:
  \url{http://dx.doi.org/10.1109/72.80202}
\BIBentrySTDinterwordspacing

\bibitem{ioannou}
M.~M. Polycarpou and P.~A. Ioannou, ``Identification and control of nonlinear
  systems using neural network models: Design and stability analysis,''
  Electrical Engineering—Systems Rep, Tech. Rep., 1991.

\bibitem{4Huang2005}
G.-B. Huang, Q.-Y. Zhu, and C.-K. Siew, ``Extreme learning machine: Theory and
  applications,'' \emph{Neurocomputing}, vol.~70, pp. 489--501, 2006.

\bibitem{huang12}
G.-B. Huang, H.~Zhou, X.~Ding, and R.~Zhang, ``Extreme learning machine for
  regression and multiclass classification.'' \emph{IEEE Transactions on
  Systems, Man, and Cybernetics, Part B}, vol.~42, no.~2, pp. 513--529, 2012.

\bibitem{oselm}
N.~ying Liang, G.~bin Huang, S.~Member, P.~Saratch, S.~Member, and
  N.~Sundararajan, ``A fast and accurate online sequential learning algorithm
  for feedforward networks,'' \emph{{IEEE} Trans. Neural Networks}, pp.
  1411--1423, 2006.

\bibitem{Yan2000}
L.~Yan, N.~Sundararajan, and P.~Saratchandran, ``Nonlinear system
  identification using lyapunov based fully tuned dynamic rbf networks,''
  \emph{Neural Process. Lett.}, vol.~12, no.~3, pp. 291--303, 2000.

\bibitem{prbs_sysid}
R.~Nowak and B.~Van~Veen, ``Nonlinear system identification with pseudorandom
  multilevel excitation sequences,'' in \emph{Acoustics, Speech, and Signal
  Processing, 1993. ICASSP-93., 1993 IEEE International Conference on}, vol.~4,
  april 1993, pp. 456 --459 vol.4.

\bibitem{Vapnik}
V.~N. Vapnik, \emph{The nature of statistical learning theory}.\hskip 1em plus
  0.5em minus 0.4em\relax New York, NY, USA: Springer-Verlag New York, Inc.,
  1995.

\end{thebibliography}

\end{document}